# Fractal Conductance Fluctuations in Generic Chaotic Cavities

Roland Ketzmerick

*Physics Department, University of California, Santa Barbara, CA 93106, USA;*
*Institut für Theoretische Physik und SFB Nichtlineare Dynamik,*
*Universität Frankfurt, D-60054 Frankfurt/Main, Germany* *

It is shown that conductance fluctuations due to phase coherent ballistic transport through a chaotic cavity *generically* are fractals. The graph of conductance vs. externally changed parameter, e.g. magnetic field, is a fractal with dimension $D = 2 - \beta/2$ between 1 and 2. It is governed by the exponent $\beta$ ($\leq 2$) of the power law distribution $P(t) \sim t^{-\beta}$ for a classically chaotic trajectory to stay in the cavity up to time $t$, which is typical for chaotic systems with a mixed (chaotic and regular) phase space. The phenomenon should be observable in semiconductor nanostructures and microwave billiards.

Phase coherent phenomena have received considerable attention over the past years. In *disordered* conductors smaller than the phase coherence length universal conductance fluctuations and weak localization have been found [1]. Recent experimental and theoretical work put emphasis on *ballistic* transport in nanostructures on semiconductor heterojunctions with chaotic classical dynamics [2–12]. Regarding conductance fluctuations the focus has been on the *specific* case of hyperbolic systems where the escape from a cavity typically is exponentially fast and spectral as well as correlation properties of the fluctuations have been predicted [13,14] and observed [15]. Phase coherent phenomena in the *generic* case of systems with a mixed (chaotic and regular) classical phase space, however, have apart from Ref. [16] received much less attention. There the chaotic part of the phase space has a very different long time behaviour, namely the escape from such a cavity is much slower and follows a *power law* similar to those reported in various chaotic model systems [17–19]. This is believed to be due to an infinite hierarchy of cantori, i.e. sets of measure zero which are partial barriers for transport, spreaded in the hierarchical structure of phase space [20,21]. Therefore one may wonder what are the fingerprints of the hierarchical phase space structure of chaotic systems and their corresponding power law escapes of chaotic trajectories on conductance fluctuations?

In the present paper this question is answered by showing that in the *generic* case (i.e. a mixed phase space with power law escape from the cavity) phase coherent ballistic transport leads to *fractal* fluctuations of the conductance, as illustrated in Fig. 1. Quantitatively, the fractal dimension $D$ of the graph of conductance vs. externally changed parameter, e.g. gate voltage or magnetic field, is shown to be given by

$$D = 2 - \beta/2 \qquad (1)$$

if the classical probability $P(t)$ of chaotic trajectories to stay longer than a time $t$ in the cavity decays like a power law

$$P(t) \sim t^{-\beta} \qquad (2)$$

with exponent $\beta \leq 2$.

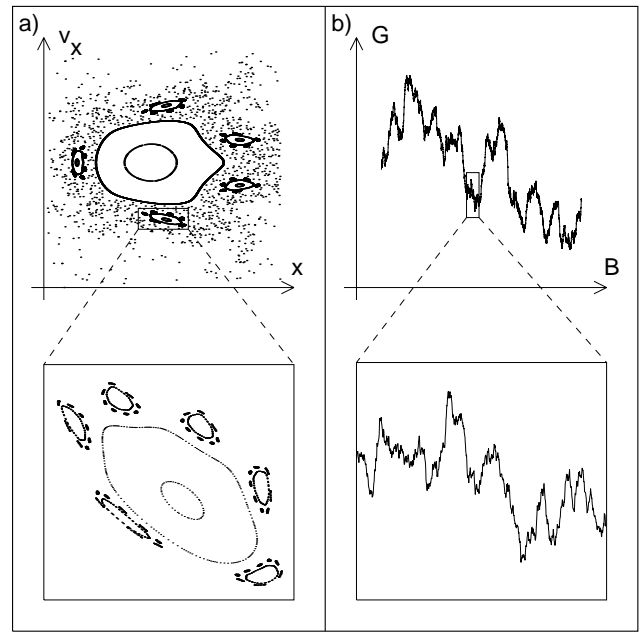

FIG. 1. (a) The hierarchical phase space structure of a 2D chaotic system and (b) fractal conductance fluctuations are shown. Their relation for phase coherent ballistic transport is derived in the text (Eqs. (9, 12)). The Poincaré surface of section shows the intersections of one chaotic and 6 regular trajectories with the $(y = 0)$-plane ($v_y > 0$) for the example of an antidot array (Ref. [4]). The conductance fluctuations are described by fractional Brownian motion according to Eqs. (10,12), where $D = 1.4$ was used for this figure.

Regular and metastable trajectories also can give rise to a power law distribution [22,12,23], however, typically with an exponent $\beta > 2$. As they group in families and their contribution to the phase coherent conductance is very subtle (see Ref. [12] for a conjecture), their contribution will not be discussed in the remainder of the paper, but rather the contribution of the chaotic part of the mixed phase space. Also, sharp resonances due to quantum tunneling between the chaotic and regular parts of phase space, as observed by Seba [24], will be neglected.

The surprising phenomenon of fractal fluctuations should be observable in semiconductor nanostructures [2,3,8,15] as well as in microwave billiards [25]. In particular such an experiment would serve as the first quantitative observation of the hierarchical phase space structure of a chaotic system. A possible experimental realization will be discussed at the end of this paper.

The two-probe conductance $G$ of a nanostructure is proportional to the sum of transmission amplitudes squared from mode $m$ of one lead to mode $n$ of the other lead

$$G = \frac{e^2}{h} \sum_{n,m} |t_{nm}|^2, \quad (3)$$

which in the semiclassical approximation are given by [14]

$$t_{nm} = \sum_s \sqrt{p_s}\, e^{\frac{i}{\hbar} S_s - i\frac{\pi}{2}\nu_s}. \quad (4)$$

Here $S_s$ and $\nu_s$ are the classical action and Maslov index of path $s$ traversing the cavity with classical transmission probability $p_s$ to go from mode $m$ to mode $n$. A small change in magnetic field or in energy will change the classical action of path $s$

$$S_s(E+\Delta E, B+\Delta B) = S_s(E,B) + t_s \Delta E + \frac{h \theta_s \Delta B}{\phi_o}, \quad (5)$$

where $t_s$ is the time for path $s$, $\theta_s = \frac{1}{2\pi}\int_s \frac{\partial \mathbf{A}}{\partial \mathbf{B}} dl$, $\nabla \times \mathbf{A} = \mathbf{B}$, and $\phi_o = h/e$ is the magnetic flux quantum. For a closed path $\theta_s$ gives the (accumulated) area enclosed by the orbit.

In analogy to previous derivations of the correlation function [16,12] from the above formulae the change in the conductance for a small change in the external parameter, e.g. the magnetic field, is found to be

$$\Delta G = G(E, B+\Delta B) - G(E,B)$$
$$= \frac{e^2}{h} \sum_{n,m} \sum_{s,u} \sqrt{p_s p_u} (e^{2\pi i(\theta_s - \theta_u)\frac{\Delta B}{\phi_o}} - 1)$$
$$\times e^{\frac{i}{\hbar}(S_s - S_u) - i\frac{\pi}{2}(\nu_s - \nu_u)} \quad (6)$$

In the semiclassical limit, where the classical actions $S_s$ and $S_u$ are much larger than $\hbar$ and can be assumed to be independent for different chaotic trajectories, the last exponential factor in Eq. (6) can be considered as a complex random number with mean 0 and variance 1 (except for $s = u$, but in that case the second but last factor vanishes). Assuming statistical independence of the actions $S_s$ for paths connecting different modes (diagonal approximation) and with the help of the central limit theorem it follows that the increment $\Delta G$ is a Gaussian distributed random variable with mean zero and variance

$$<(\Delta G)^2> = \left(\frac{e^2}{h}\right)^2 \sum_{n,m}\sum_{s,u} p_s p_u \left|e^{2\pi i(\theta_s - \theta_u)\frac{\Delta B}{\phi_o}} - 1\right|^2$$
$$= 2\left(\frac{e^2}{h}\right)^2 \sum_{n,m}\left(\left(\sum_s p_s\right)^2 - \left|\sum_s p_s e^{2\pi i \frac{\theta_s \Delta B}{\phi_o}}\right|^2\right). \quad (7)$$

Replacing the sum over paths $s$ by an integral over the distribution $\mathrm{d}P(\theta)/\mathrm{d}\theta$ of areas $\theta$ enclosed by classical paths, assuming $P(\theta)$ to be independent of the modes $n$ and $m$, and by restricting to positive areas $\theta \geq 0$ for simplicity one finds

$$<(\Delta G)^2> \sim 1 - \left|\int_0^\infty \mathrm{d}\theta \frac{\mathrm{d}P(\theta)}{\mathrm{d}\theta} e^{2\pi i \frac{\theta \Delta B}{\phi_o}}\right|^2. \quad (8)$$

Power law distributions of the enclosed areas

$$P(\theta) \sim \theta^{-\gamma} \quad (9)$$

for large $\theta$ then cause the variance of the increments to scale like

$$<(\Delta G)^2> \sim (\Delta B)^\gamma \quad ; \quad \gamma \leq 2 \quad (10)$$
$$<(\Delta G)^2> \sim (\Delta B)^2 \quad ; \quad \gamma > 2 \quad (11)$$

under small changes in the magnetic field. Thus for $\gamma \leq 2$ the graph of conductance vs. magnetic field has the same statistical properties as a Gaussian random process with increments of mean zero and variance $(\Delta B)^\gamma$. Such processes are known as fractional Brownian motion [26] and have the property that their graph is a fractal with dimension

$$D = 2 - \gamma/2 \quad (12)$$

between 1 and 2 (Fig. 1b). Similarly, for small changes of the energy e.g. by changing a gate voltage and for a power law distribution $P(t) \sim t^{-\beta}$ ($\beta \leq 2$) of classical sojourning times in the cavity larger than $t$ the fractal dimension is given by $D = 2 - \beta/2$.

Thus the graph of conductance vs. magnetic field (energy) is a reproducible fractal line, whenever the classical probability $P(\theta)$ ($P(t)$) decays like a power law with an exponent smaller than 2. This phenomenon, which we call *fractal conductance fluctuations* is generic for ballistic nanostructures, as in a typical system one has a mixed phase space with a hierarchical phase space structure (Fig. 1a) and a hierarchy of cantori causing the power law of the staying probability [20,21]. In the remainder of the paper some general remarks regarding the observability of fractal conductance fluctuations are given and an experimental realization is proposed.

Equation (4) makes use of the semiclassical approximation for the transmission amplitudes which is valid up to some finite time $t^*$ only. Therefore the fine scale fluctuations of the conductance due to trajectories staying longer than this time $t^*$ in the cavity will be washed out by quantum mechanics. Thus the graph of conductance vs. externally changed parameter will look like a fractal only on scales larger than

$$\Delta E^* \sim \hbar/t^* \text{ or } \Delta B^* \sim \phi_o/\theta(t^*). \tag{13}$$

As for any other fractal in the physical world there is not only such a smallest, but also a largest scale. Here it is determined by the time and the area of the onset of the power law behaviour of $P(t)$ and $P(\theta)$, respectively.

How big the range of validity for the semiclassical approximation is and how it depends on the effective $\hbar$ of the system, is a subject of current research [27]. A rough estimate for the present situation may be that only those parts of the hierarchical phase space structure can be resolved that are larger than $\hbar$. This will lead to a breakdown of the power law staying probability at some time $t^* \sim (\hbar/S)^{-\alpha}$ with $\alpha$ depending on the system and $S$ being the action of a typical trajectory.

What is the effect of disorder? As long as the *elastic* scattering can be modelled by smooth impurity potentials it will not change the generic properties of the mixed phase space. Thus fractal conductance fluctuations may well be observable in the presence of disorder. For strong disorder, however, the fractal dimension $D$ may depend on the specific disorder configuration. The situation would be different for integrable systems and some idealized chaotic systems with hard walls, e.g. the stadium billiard, where power law distributions have been found [22,12,23]. The latter are due to families of metastable orbits and would break down even for small disorder. *Inelastic* scattering, e.g. due to electron-electron interaction, destroys the phase coherence and therefore the fractal conductance fluctuations on small scales. This scale is determined by the inelastic scattering time $t_\varphi$ in the same way as in Eq. (13) for $t^*$.

How severe is the restriction to systems with $\beta, \gamma \leq 2$? Exponents $\beta < 2$ as well as $\beta > 2$ have been reported in various systems and parameter ranges [17–19,21,16], so far, however, it is not possible to predict the exponent $\beta, \gamma$ for a given system. For the case $\beta, \gamma \geq 2$ it follows from Eq. (11) that the graph of conductance vs. externally changed parameter is a smooth line with dimension 1. Thus not every such chaotic system gives rise to fractal conductance fluctuations with $D > 1$.

To test the above semiclassical treatment of the conductance numerically by comparison with a quantum mechanical calculation, e.g. a recursive discrete Greens function method [28], is in general beyond current computational resources, as an approximating discrete lattice must be fine enough to resolve the hierarchical structure in classical phase space on at least a few levels. In a one-dimensional periodic array of potential wells studied by Lai et al. [16], however, it was possible to confirm a semiclassically derived cusp in the correlation function $C(\Delta E) = C(0) - (\Delta E)^\delta$ for $P(t) \sim t^{-\delta}$. As the correlation function is related to the variance of the increments by $C(\Delta E) = C(0) - \frac{1}{2} < (\Delta G)^2 >$ their calculations do confirm the semiclassical derivation of the variance $< (\Delta G)^2 >$ (Eq. (10)), from which the occurrence of fractal conductance fluctuations is concluded.

Since ballistic transport in chaotic cavities has been studied successfully in GaAs/AlGaAs heterojunctions [2,3,6–8,11,15] these should be suitable for the observation of fractal conductance fluctuations. The lower left inset of Fig. 2 shows an example for a geometry, namely a widened channel with parabolic walls, that gives rise to a mixed classical phase space. In a perpendicular magnetic field many trajectories will be trapped close to regular cyclotron-like orbits for some time (upper right inset of Fig. 2). This gives rise to power law distributions for the staying time $P(t) \sim t^{-1.7}$ and for the area enclosed by a trajectory $P(\theta) \sim \theta^{-1.7}$ (Fig. 2). For this cavity Eq. (12) predicts fractal conductance fluctuations with a dimension $D = 1.15$. Similar results are found for cavities with other parameters or with an antidot added to the center of the cavity. The power law exponents range from 0.9 to 3.5 corresponding to fractal conductance fluctuations with $1 \leq D \leq 1.55$. In all these examples with magnetic fields $B \neq 0$ the power law exponents of $P(t)$ and $P(\theta)$ turn out to be identical.

The energy and the magnetic field scale for fluctuations due to trajectories making $N$ revolutions is given by (Eqs. (4) and (5))

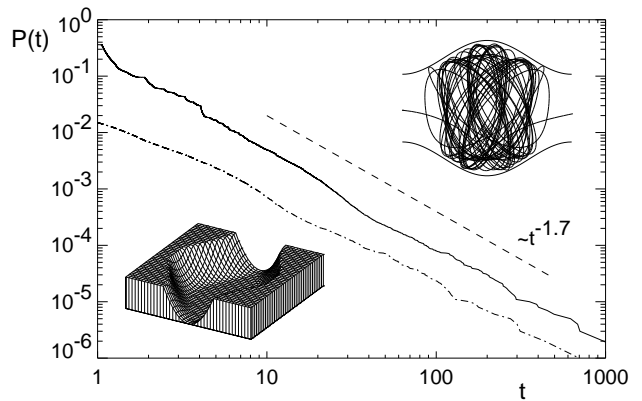

FIG. 2. The probability $P(t)$ to stay in the cavity (lower left inset) for a time greater than $t$ in units of $t_o$ in a perpendicular magnetic field $B = 0.6 B_o$ decays with a power law $P(t) \sim t^{-1.7}$ (solid line). The probability $P(\theta)$ for enclosed areas greater than $\theta$ in units of $\theta_o$ follows the same power law behaviour (dash-dotted line). The upper right inset shows a typical trajectory trapped in the hierarchical phase space structure giving rise to the power law behaviour. For this figure a cavity formed by two cosines with maximum distance $d$, minimum distance $d/2$, length $5d/4$, and parabolic walls of width $d/4$ was used.

$$\Delta E_N = \frac{h}{t_o N} \text{ and } \Delta B_N = \frac{\phi_o}{\theta_o N} \quad (14)$$

where $t_o$ and $\theta_o$ are the time and the area for one revolution. For a free cyclotron orbit of diameter $d$ these are given by $t_o = \pi d/v_F$ and $\theta_o = \pi d^2/4$ at magnetic field $B_o = 2\pi m^*/(e t_o)$ with Fermi velocity $v_F = \hbar\sqrt{2\pi n}/m^*$, effective mass $m^*$, and the 2D electron density $n$. This gives rise to an effective Planck's constant $\hbar_{\text{eff}} = \hbar/\int p\,dq = 1/(2^{1/2}\pi^{3/2}dn^{1/2})$.

In order to observe fractal conductance fluctuations over at least one order of magnitude one needs $N \gtrsim 10$ revolutions and thus a phase coherence length $\lambda_\varphi \gtrsim 10\pi d$. In addition the validity of the semiclassical approximation over this length scale is required. As mentioned above, the latter is not known exactly. On the other hand in an antidot array of period $300nm$ magnetotransport anomalies due to classical electron trajectories surrounding 21 antidots have been observed [2] corresponding to a length of about $5\mu m$. The phase coherence length in these 2D electron systems is about $10\mu m$ [6]. Therefore a size of the cavity of about $d = 300nm$ would be favourable for experiments with the current technology. For smaller $d$ the scale of the fluctuations $\Delta B_N$ would be too large for the assumption implicit in the above semiclassical transport theory (Eq. (4)) that on the scale of the fluctuations only the phases but not the classical trajectories themselves are changed. For the example shown in Fig. 2 the exponent $\beta$ is unchanged [29] on a magnetic field range $\Delta B_\beta \approx 0.1 B_o$. For $d = 300nm$, $n = 5 \times 10^{15}m^{-2}$, and $m^* = 0.067 m_e$ one finds $\Delta B_\beta \approx 80mT$, which must be compared with $\Delta B_N \approx 60mT/N$. Thus fractal conductance fluctuations should be observable on one order of magnitude in these experiments with $d \approx 300nm$, which provides a lower bound. More and more details of these fractal fluctuations will be observable for samples with larger phase coherence length.

The derived relation between a classical power law staying probability and the fractal dimension of the conductance fluctuations is of semiclassical origin. Therefore, although at first sight very similar, it is unrelated to the general relation of the quantum mechanical power law staying probability to a multifractal dimension of the local density of states [30] which in some examples is also related to a multifractal dimension of the eigenfunctions [31].

In conclusion, fractal conductance fluctuations are a generic phenomenon of phase coherent ballistic transport. They are due to the hierarchical phase space structure of mixed chaotic systems and should be observable in nanostructures on semiconductor heterojunctions.

It is a pleasure to thank Walter Kohn for discussions on conductance fluctuations that stimulated this work and for his hospitality during a stay at UCSB. Further helpful discussion with Theo Geisel are gratefully acknowledged.

This work was supported by the Deutsche Forschungsgemeinschaft and partially by the NSF under DMR93-08011.

* Present address.
[1] See e.g. B. L. Altshuler, P. A. Lee, and R. A. Webb (eds.), *Mesoscopic Phenomena in Solids* (North-Holland, New York, 1991).
[2] D. Weiss, M. L. Roukes, A. Menschig, P. Grambow, K. v. Klitzing, and G. Weimann, Phys. Rev. Lett. **66**, 2790 (1991).
[3] R. Schuster, K. Ensslin, J. P. Kotthaus, M. Holland, and S. P. Beaumont, Superlattices and Microstructures **12**, 93 (1992).
[4] R. Fleischmann, T. Geisel, and R. Ketzmerick, Phys. Rev. Lett. **68**, 1367 (1992); Europhys. Lett. **25**, 219 (1994).
[5] T. Geisel, R. Ketzmerick, and O. Schedletzky, Phys. Rev. Lett. **69**, 1680 (1992).
[6] D. Weiss, K. Richter, A. Menschig, R. Bergmann, H. Schweizer, K. von Klitzing, and G. Weimann, Phys. Rev. Lett **70**, 4118 (1993).
[7] R. Schuster, K. Ensslin, D. Wharam, S. Kühn, J. P. Kotthaus, G. Böhm, W. Klein, G. Tränkle, and G. Weimann, Phys. Rev. B **49**, 8510 (1994).
[8] A. S. Sachrajda, Y. Feng, R. P. Taylor, G. Kirczenow, L. Henning, J. Wang, P. Zawadzki, and P. T. Coleridge, Phys. Rev. B **50**, 10856 (1994).
[9] G. Hackenbroich and F. v. Oppen, Europhys. Lett. **29**, 151 (1995).
[10] K. Richter, Europhys. Lett. **29**, 7 (1995).
[11] C. M. Marcus, R. M. Westervelt, P. F. Hopkins, and A. C. Gossard, Chaos **3**, 643 (1993).
[12] H. U. Baranger, R. A. Jalabert, and A. D. Stone, Chaos **3**, 665 (1993).
[13] R. Blümel and U. Smilansky, Phys. Rev. Lett **60**, 477 (1988); E. Doron, U. Smilansky, and A. Frenkel, Physica D **50**, 367 (1991).
[14] R. A. Jalabert, H. U. Baranger, and A. D. Stone, Phys. Rev. Lett **65**, 2442 (1990).
[15] C. M. Marcus, A. J. Rimberg, R. M. Westervelt, P. F. Hopkins, and A. C. Gossard Phys. Rev. Lett **69**, 506 (1992).
[16] Y.-C. Lai, R. Blümel, E. Ott, and C. Grebogi, Phys. Rev. Lett **68**, 3491 (1992).
[17] C. F. F. Karney, Physica **8** D, 360 (1983).
[18] B. V. Chirikov, D. L. Shepelyanski, Physica **13** D, 395 (1984).
[19] P. Grassberger and H. Kantz, Phys. Lett. **113** A,167 (1985).
[20] J. D. Meiss and E. Ott, Physica **20** D, 387 (1986).
[21] T. Geisel, A. Zacherl, and G. Radons, Phys. Rev. Lett. **59**, 2503 (1987); Z. Phys. B **71**, 117 (1988).
[22] W. Bauer and G. F. Bertsch, Phys. Rev. Lett. **65**, 2213 (1990).

[23] H. Alt, A.-D. Gräf, H. L. Harney, R. Hofferbert, H. Rehfeld, A. Richter, and P. Schardt, Phys. Rev. E **53**, 2217 (1996).
[24] P. Seba, Phys. Rev. E **47**, 3870 (1992).
[25] J. Stein and H.-J. Stöckmann, Phys. Rev. Lett. **68**, 2867 (1992); H.-D. Gräf, H. L. Harney, H. Lengeler, C. H. Lewenkopf, C. Rangacharyulu, A. Richter, P. Schardt, and H. A. Weidenmüller, Phys. Rev. Lett. **69**, 1296 (1992).
[26] B. B. Mandelbrot, *The Fractal Geometry of Nature* (Freeman, San Francisco, 1982).
[27] M. Sepúlveda, S. Tomsovic, and E. J. Heller Phys. Rev. Lett. **69**, 402 (1992) and reference [7] in [16].
[28] H. U. Baranger, D. P. DiVincenzo, R. A. Jalabert, and A. D. Stone, Phys. Rev. B **44**, 10637 (1991).
[29] When varying a paramter the power law exponent $\beta$ might fluctuate for very large times, as described in: Y.-C. Lai, M. Ding, C. Grebogi, and R. Blümel, Phys. Rev. A **46**, 4661 (1992). For the finite range of times $t < t_\varphi$ relevant for a physical sample, however, the power law exponent $\beta$ is found to be much more stable for the examples studied in this paper.
[30] R. Ketzmerick, G. Petschel, and T. Geisel, Phys. Rev. Lett. **69**, 695 (1992).
[31] B. Huckestein and L. Schweitzer, Phys. Rev. Lett. **72**, 713 (1994).